\documentclass[twocolomn,fleqn]{pasj00}
\usepackage{longtable}
\usepackage{lscape}

\draft

\def\commenta{$^*$}
\def\commentb{$^\dagger$}
\def\commentc{$^\ddagger$}
\def\commentd{$^\S$}

\newcounter{author}
\def\authorcount#1#2{\refstepcounter{author}\label{#1}
                     \altaffiltext{\ref{#1}}{#2}}

\begin{document}
\SetRunningHead{Pavlenko et al.}{NY Serpentis: Diversity of Normal Outbursts}

\Received{201X/XX/XX}
\Accepted{201X/XX/XX}

\title{NY Serpentis: SU UMa-Type Nova in the Period Gap with Diversity of
Normal Outbursts}

\author{Elena~P.~\textsc{Pavlenko},\altaffilmark{\ref{affil:CrAO}*}$^,$\altaffilmark{\ref{affil:Kyoto}}
        Taichi~\textsc{Kato},\altaffilmark{\ref{affil:Kyoto}}
        Oksana~I.~\textsc{Antonyuk},\altaffilmark{\ref{affil:CrAO}}
        Tomohito~\textsc{Ohshima},\altaffilmark{\ref{affil:Kyoto}}
        Franz-Josef~\textsc{Hambsch},\altaffilmark{\ref{affil:GEOS}}$^,$\altaffilmark{\ref{affil:BAV}}$^,$\altaffilmark{\ref{affil:Ham}}
        Kirill~A.~\textsc{Antonyuk},\altaffilmark{\ref{affil:CrAO}}
        Aleksei~A.~\textsc{Sosnovskij},\altaffilmark{\ref{affil:CrAO}}
				Alex~V.~\textsc{Baklanov},\altaffilmark{\ref{affil:CrAO}}
        Sergey~Yu.~\textsc{Shugarov},\altaffilmark{\ref{affil:Sternberg}}$^,$\altaffilmark{\ref{affil:Slovak}}
        Nikolaj~V.~\textsc{Pit},\altaffilmark{\ref{affil:CrAO}}
        Chikako~\textsc{Nakata},\altaffilmark{\ref{affil:Kyoto}}
        Gianluca~\textsc{Masi},\altaffilmark{\ref{affil:Masi}}
        Kazuhiro~\textsc{Nakajima},\altaffilmark{\ref{affil:Njh}}
        Hiroyuki~\textsc{Maehara},\altaffilmark{\ref{affil:Kiso}}
        Pavol~A.~\textsc{Dubovsky},\altaffilmark{\ref{affil:Dubovsky}}
        Igor~\textsc{Kudzej},\altaffilmark{\ref{affil:Dubovsky}}
        Maksim~V.~\textsc{Andreev},\altaffilmark{\ref{affil:Terskol}}$^,$\altaffilmark{\ref{affil:ICUkraine}}
        Yuliana~G.~\textsc{Kuznyetsova},\altaffilmark{\ref{affil:MAO}}

        Kirill~A.~\textsc{Vasiliskov},\altaffilmark{\ref{affil:Kyiv}}
}

\authorcount{affil:CrAO}{
     Crimean Astrophysical Observatory, 98409, Nauchny 19/17, Crimea, Ukraine}

\authorcount{affil:Kyoto}{
     Department of Astronomy, Kyoto University, Kyoto 606-8502}
\email{$^*$eppavlenko@gmail.com}

\authorcount{affil:GEOS}{Groupe Europ\'een d'Observations Stellaires (GEOS),
     23 Parc de Levesville, 28300 Bailleau l'Ev\^eque, France
     }

\authorcount{affil:BAV}{Bundesdeutsche Arbeitsgemeinschaft f\"ur Ver\"anderliche
Sterne (BAV), Munsterdamm 90, 12169 Berlin, Germany
     }

\authorcount{affil:Ham}{Vereniging Voor Sterrenkunde (VVS), Oude Bleken 12,
2400 Mol, Belgium
     }

\authorcount{affil:Sternberg}{
     Sternberg Astronomical Institute, Lomonosov Moscow University,
     Universitetsky Ave., 13, Moscow 119992, Russia}

\authorcount{affil:Slovak}{
     Astronomical Institute of the Slovak Academy of Sciences, 05960,
     Tatranska Lomnica, the Slovak Republic}

\authorcount{affil:Masi}{
     The Virtual Telescope Project, Via Madonna del Loco 47, 03023
     Ceccano (FR), Italy}

\authorcount{affil:Njh}{
     Variable Star Observers League in Japan (VSOLJ),
     124 Isatotyo, Teradani, Kumano, Mie 519-4673}

\authorcount{affil:Kiso}{
     Kiso Observatory, Institute of Astronomy, School of Science,
     The University of Tokyo 10762-30, Mitake, Kiso-machi, Kiso-gun,
     Nagano 397-0101}

\authorcount{affil:Dubovsky}{
     Vihorlat Observatory, Mierova 4, Humenne, Slovakia}

\authorcount{affil:Terskol}{
     Institute of Astronomy, Russian Academy of Sciences, 361605 Peak Terskol,
     Kabardino-Balkaria, Russia}

\authorcount{affil:ICUkraine}{
     International Center for Astronomical, Medical and Ecological Research
     of NASU, Ukraine 27 Akademika Zabolotnoho Str. 03680 Kyiv,
     Ukraine}

\authorcount{affil:MAO}{
     Main Astronomical Observatory
     of NASU, Ukraine 27 Akademika Zabolotnoho Str. 03680 Kyiv,
     Ukraine}

\authorcount{affil:Kyiv}{
    Taras Shevchenko
    National University, Kyiv 022, prosp. Glushkova 2, Ukraine}



\maketitle

\begin{abstract}
We present photometric study of NY Ser, an in-the-gap SU UMa-type
nova, in 2002 and 2013.  We determined the duration
of the superoutburst and the mean superhump period to be 18~d
and 0.10458~d, respectively.
We detected in 2013 that NY Ser showed two distinct states
separated by the superoutburst.  A state of rather infrequent
normal outbursts lasted at least 44~d before the superoutburst
and a state of frequent outbursts started immediately after
the superoutburst and lasted at least for 34~d.
Unlike a typical SU UMa star with bimodal distribution of
the outbursts duration,
NY Ser displayed a diversity of normal outbursts.
In the state of infrequent outbursts, we detected
a wide $\sim$12~d outburst accompanied by 0.098~d orbital modulation
but without superhumps ever established in NY Ser.
We classified this as the ``wide normal outburst".
The orbital period dominated both in quiescence and during
normal outbursts in this state.  In the state
of the most frequent normal outbursts, the 0.10465~d
positive superhumps dominated and co-existed with the orbital modulation.
In 2002 we detected the normal outburst of ``intermediate"
5--6~d duration that was also accompanied by orbital modulations.
\end{abstract}

\section{Introduction}

Cataclysmic variables (CVs) are close binary stars at the late stage of
their evolution (\cite{war95suuma}; \cite{hel01book} for reviews).
In these binaries the late-type star fills in its Roche lobe and
transfers matter onto the compact component (white dwarf)
through an accretion disk.
The thermal-viscous instability in the accretion disk causes outbursts
(\cite{can93DIreview}; \cite{las01DIDNXT}).

The SU UMa-type stars are a subgroup of the short-periodic
non-magnetic cataclysmic variables occupying a region of
the orbital periods bounded by the minimum period
at $\sim$76 min and the upper limit of the ``period gap", 3.18~hr
(\cite{kni06cvsecondary}).
SU UMa stars show the well-known bimodality of the outburst
duration (\cite{vanpar83superoutburst}; \cite{war95suuma}):
the narrow ``normal" outbursts lasting for a few days and
the wide superoutbursts that are longer in duration by a factor of
5--10 in the same system.  Typical superoutburst lasts for
about two weeks.
Several normal outbursts occur between the superoutbursts.
The amplitudes of superoutbursts are slightly higher than
or equal to those of normal outbursts.
\citet{vanpar83superoutburst} pointed out that almost
all dwarf novae have a bimodal distribution of
the outburst duration.  Only SU UMa stars, however, have
the wide outbursts (superoutbursts) accompanied by
what are called positive superhumps, which are brightness variations
with periods a few percent longer than the orbital period.

According to the modern paradigm of the tidal instability,
the 3:1 resonance in the accretion disk orbiting the white dwarf
in systems with mass ratio $q = m_2/m_1\leq0.25$,
where $m_2$ is the mass of the late type star and $m_1$ is that
of the white dwarf, is responsible for its eccentric deformation,
resulting in positive superhumps
[\citet{whi88tidal}, \citet{hir90SHexcess}, \citet{lub91SHa}].
\citet{osa96review} proposed a thermal-tidal instability model
in which the ordinary thermal instability
is coupled with the tidal instability.

The detailed study of the superoutbursts and evolution of
the positive superhumps is given in
the series of papers [\citet{Pdot}; \citet{Pdot2};
\citet{Pdot3}; \citet{Pdot4}; \citet{Pdot5}; \citet{Pdot6}].
The durations of superoutburst vary from one system to another,
but is rather stable value for the same star
(\cite{war95suuma}) in different occasions.\footnote{There are,
however, exceptions especially in systems with low outburst
frequencies.  The notable example is BC UMa \citep{rom64bcuma}.
See also \citet{kat95wxcet}}.

As for the durations of normal outbursts, they depend both
on the orbital period and on the supercycle phase.
It was found by the early investigations \citep{vanpar83superoutburst}
that the duration of the narrow outburst $t_b$ for several CVs
increases with the orbital period.  For the SU UMa stars,
the empirical correlation between $t_b$ and the orbital period
was $t_b \sim$2.5--5~d.  Recently \citet{can12v344lyr}
explored the Kepler light curves of the SU UMa stars V1504 Cyg
and V344 Lyr and found a systematic increase of $t_b$ between
two consecutive superoutbursts, that were $t_b$=1.1--2.9~d for
V1504 Cyg and $t_b$=2.5--5.0~d for V344 Lyr.

The 2.15--3.18~hr period gap \citep{kni06cvsecondary}
represents as division between the long-period systems with
high mass-transfer rates (above the period gap) and
short-period systems with low-mass transfer rates
(below the period gap).
The evolution of the long-period cataclysmic variables is
mainly driven by magnetic braking,
while the evolution of the short-period
ones is driven by gravitational wave radiation
[\citet{kra62ugem}, \citet{ver81magneticbraking},
\citet{kni06cvsecondary}].
According to the theoretical predictions, the secondary loses
contact with its Roche lobe at an orbital period
$\sim$3~hr and the the secondary reaches contact again at
an orbital period $\sim$2~hr.  This causes the dearth
of the CVs in this region of orbital periods,
so this region is called the ~"period gap".
Despite that a number of SU UMa-type stars in the period gap
have grown with time [see, for example, \citet{kat03CVperiodgap};
\citet{dai12CVgap}; \citet{sch06admen}], the properties of
these systems are still poorly studied.

NY Ser is the first in-the-period gap SU UMa-type dwarf nova.
It was discovered as an ultraviolet-excess object PG 1510$+$234
(\cite{PGsurvey}; \cite{gre82PGsurveyCV}).  \citet{iid95nyser}
identified that this object is a frequently outbursting
dwarf nova.  \citet{nog98nyser} detected superhumps during
a long outburst in 1996 April, establishing the SU UMa-type
classification. Its supercycle was estimated as 85--100~d
in 1996 by \citet{nog98nyser} and 60--90~d according to
\citet{pat03suumas}, who observed mostly in 1999.  NY Ser
displayed rather frequent normal outbursts every
6--9~d [\citet{nog98nyser}; \citet{iid95nyser}] of
durations about 3~d that are typical for the most of
SU UMa-type dwarf novae.
However the outburst activity of NY Ser in recent years
differed from those of ``normal'' SU UMa-type stars.
Using the AAVSO data, \citet{Pdot5} found the existence of
the outbursts with ``intermediate" durations (about 4~d)
that were observed in 2011--2012 in addition to
normal outbursts with durations of $\sim$3~d and
superoutburst with longer durations.

The period of positive superhump of NY Ser has been estimated by
both \citet{nog98nyser} and \citet{pat03suumas}.
However, the details of the period evolution were
not defined due to the insufficient amount of data.
Only 3-d time series was obtained by
\citet{nog98nyser} during the 1996 superoutburst and
the 5-d one was obtained by \citet{pat03suumas} during
the 1999 superoutburst.
The analysis yielded the periods of 0.106~d and 0.104~d, respectively.
This discrepancy of periods is probably because the superhump period
was measured for different parts of the superoutburst
between these two measurements.

\citet{pat03suumas} found the orbital period of 0.0975~d
in quiescence and confirmed this value [0.09756(3)~d] on the
8-d baseline covering both quiescence and outburst.

We undertook the photometric study of NY Ser to clarify the
peculiarities of the outbursts and the behavior during
the different phases of outburst activity.

\section{Observations and data reduction}

The photometric CCD observations of NY Ser have been carried out
in 2002 and 2013.  In 2002, the star was observed for a period of
$\sim$66~d at the Crimean Astrophysical Observatory (CrAO)
and in 2013 for a period of $\sim$100~d
at several observatories located at different longitudes
The detailes of observations are given in the \ref{tab:log}
available in electronic form only. The standard
aperture photometry (de-biasing, dark subtraction and flat-fielding)
was used for measuring of the variable and comparison star.
The last one is the star USNO B1.0 1132-0246239.
On 2006 July 07, we measured the brightness of the comparison star
relative to the Landolt sequence stars 10971, having $V=11.493$,
$B-V=0.323$, $V-R=0.186$ and 109381, having $V=11.73$, $B-V=0.704$,
$V-R=0.428$.  We determined the magnitudes of the comparision star
to be $B=16.48$, $V=15.92$, $R=15.56$ [\citet{ski07UBVRI} gave
$V=15.89$ and $B-V=0.58$.]

Most of observations have been carried out without usage
of filters, giving a system close to the $R$ band in our case.
The accuracy of a single brightness measure depended on the telescope,
exposure time, weather condition and brightness of NY Ser.
It was 0.007--0.03 mag during outburst and superoutburst and
0.05--0.10 mag in quiescence.  The much higher accuracy about
0.005 mag during outbursts and 0.01 mag in quiescence was
achieved in observations with the 2.6-m telescope.

The times of observations are expressed in Barycentric Julian Days (BJD).
A phase dispersion minimization (PDM; \cite{PDM}) is used for
period analysis, and 1$\sigma$ errors was estimated by
the methods of \citet{fer89error} and \citet{Pdot2}.
Before starting the period analysis, we corrected zero-point of
data difference for different telescopes and subtracted
a long-term trend of the outbursts and quiescent light curve
by subtracting smoothed light curve obtained by locally-weighted
polynomial regression (LOWESS, \cite{LOWESS}).

\section{2002 Light Curve}

\begin{figure*}
\begin{center}
\FigureFile(115mm,80mm){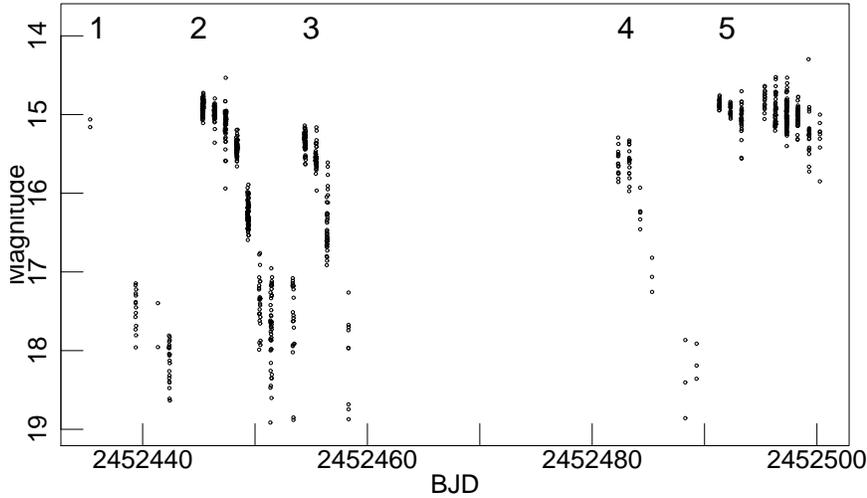}
\end{center}
\caption{Overall light curve of NY Ser in 2002.}
\label{fig:nyser2002}
\end{figure*}

The  observations of NY Ser carried out in 2002 are shown in figure
\ref{fig:nyser2002}. The 66-d light curve
covered five outbursts of different duration
that in average were separated by 9~d. While there is only indication to the first
outburst, the next two outbursts were observed in better details. The second outburst was detected at
BJD 2452445--2452450 and had duration at least 5~d (the possibility of 6~d also
cannot be excluded). The third outburst occurred around BJD 2452455 and lasted
for $\sim$ 3~d.
The last two outbursts
also were separated by $\sim$ 9~d in average. The outburst around BJD 2452485
has not been properly defined, but
its duration was not shorter that 3~d. The next outburst started at BJD 2452490
and lasted at least for 11~d. In what follows, we call the second outburst in 2002, as the
2002 first wide outburst and the fifth one as the 2002 second wide outburst.
The mean amplitude of both wide outbursts was about 2.5 mag.

\section{2013 light curve}

\begin{figure*}
\begin{center}
\FigureFile(115mm,80mm){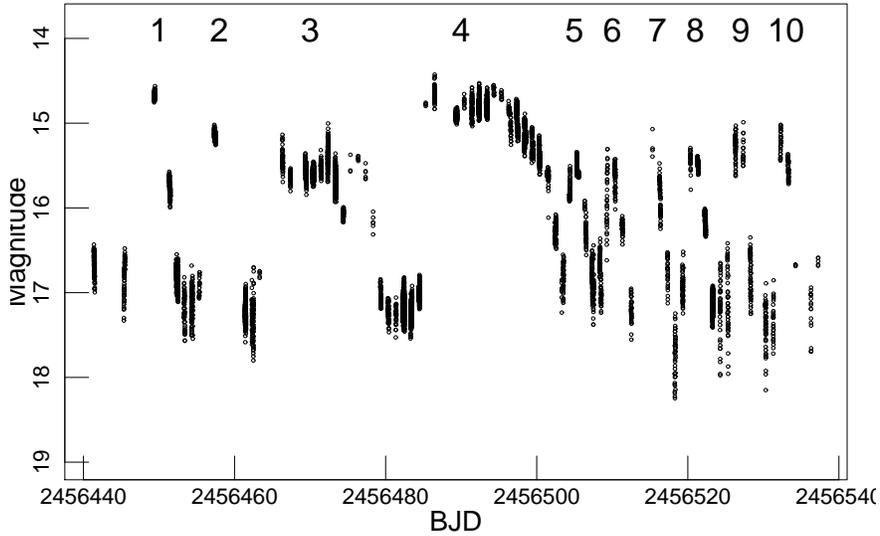}
\end{center}
\caption{Overall light curve of NY Ser in 2013.}
\label{fig:nyser2013}
\end{figure*}

The long-term light curve of NY Ser is presented in
figure \ref{fig:nyser2013}. This entire light curve
consists of ten outbursts.  Contrary to the ``usual''
SU UMa-type novae, having the sequences of short-term
normal outbursts between superoutbursts, NY Ser demonstrated
a large variety of outbursts during the 100-d interval.
During the 44-d segment before BJD 2456464, the outburst behavior
of NY Ser resembled its known
behavior (\cite{nog98nyser}; \cite{pat03suumas}):
the the first and the second outbursts in figure \ref{fig:nyser2013} looked like
the normal ones separated by $\sim$8~d and lasted for the 3--3.5~d.
Their approximate amplitudes (about 2.5 mag) are consistent
with the previously reported values.
The third outburst in figure \ref{fig:nyser2013} that appeared approximately at the
expected time was, however, entirely different.
First, its duration was no shorter than 12~d but not longer
than 15~d that never happened among the normal outbursts of
the SU UMa-type stars, but is common for the duration of
a superoutburst.  Second, the mean amplitude of
this outburst probably was less than those of the previous
normal outbursts and did not exceed 1.8 mag.
Third, the profile of this outburst looks rather symmetrical
with a structured maximum lasting for at least 11~d. We call this 
the third outburst in 2013 as the 2013 first wide outburst below.

After the 5-d minimum following this outburst,
the next wide outburst (the fourth outburst in 2013 and we call
hereafter the 2013 second wide outburst) resembling a superoutburst was detected.
Its amplitude was about 2.5 mag, the duration was 18~d,
taking into account the precursor-like structure around
BJD 2456485--2456487.  Note also the prominent round shape of
the maximum of this outburst.  The slow decline lasted only
for 6~d and occurred with a rate of $\sim$0.12 mag d$^{-1}$.

The sequence of six short outbursts with a mean amplitude
of $\sim$2~mag has been observed immediately after the end of
the longest outburst and this sequence lasted for 34~d.
Their frequency of the outbursts changed dramatically:
the separation between first two outbursts was only 5~d
(that never was recorded before), the rest of the outbursts
appeared every 6~d.  Note that the amplitudes of the first
two outbursts were also the smallest ones, reaching only 1.5 mag.

While the short outbursts have no doubt to be the ``normal"
outbursts according to the accepted definition of such outbursts
in the SU UMa stars, the nature of two wide outbursts requires
further study.  We investigated short-term nightly periodicity
(potential orbital and superhumps periods) to clarify
 the nature of these outbursts.

\section{Short-Term Variability at Different States of the Dwarf Nova Activity}

\begin{figure}
\begin{center}
\FigureFile(80mm,120mm){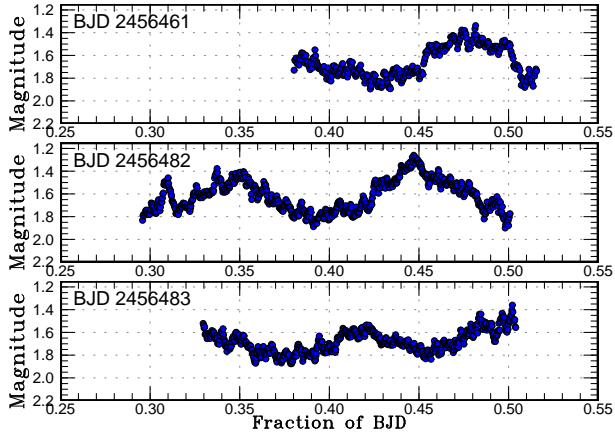}
\end{center}
\caption{Example of the light curves for selected nights in 2013 quiescent state.
The magnitudes are given
relative to the comparison star.}
\label{fig:ztsh1}
\end{figure}

\begin{figure}
\begin{center}
\FigureFile(80mm,120mm){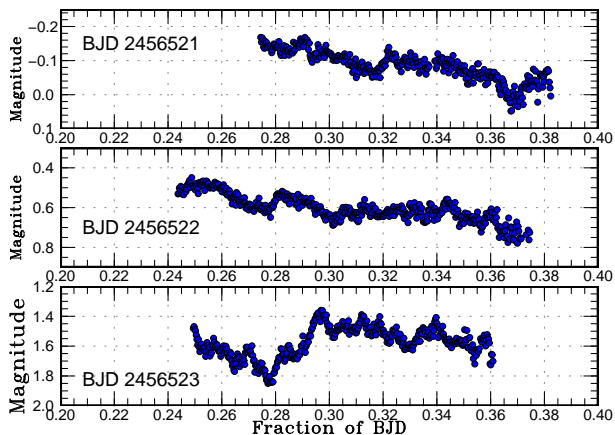}
\end{center}
\caption{Example of the light curves for the declining branch of the normal outburst.
The magnitudes are given
relative to the comparison star.}
\label{fig:ztsh2}
\end{figure}

\begin{figure}
\begin{center}
\FigureFile(80mm,120mm){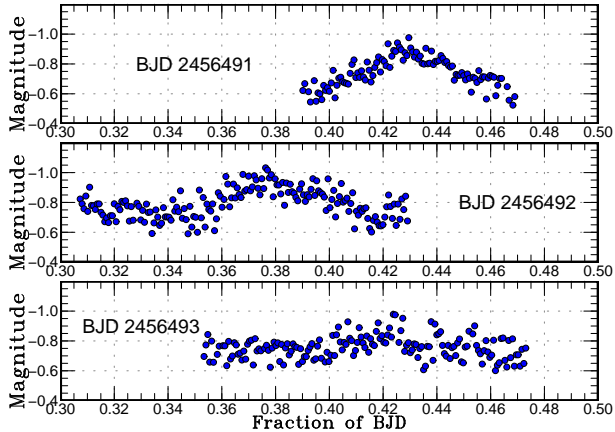}
\end{center}
\caption{Example of the nightly light curves for the 2013 superoutburst. The magnitudes are given
relative to the comparison star.}
\label{fig:superoutburst}
\end{figure}

In all states of outburst activity in NY Ser, it displayed
short-term brightness variations.  Some examples of
the nightly light curves are shown in figures \ref{fig:ztsh1},
\ref{fig:ztsh2}, \ref{fig:superoutburst}.

We studied the periodicity of NY Ser for the two data sets at
the selected states of its activity in a region of frequencies
including the orbital and superhumps frequencies.
The corresponding periodograms contain indication to
the real signal and its one-day aliases.
In the case of insufficient quantity and quality of
the data, the formal probability of the real signal not always
is higher than a false (aliased) one.  So in all the periodograms
we preferred those period as the real one that coincided or
was close to the known periods ever observed in NY Ser.

\subsection{2002 First Wide Outburst}

\begin{figure}
\begin{center}
\FigureFile(80mm,120mm){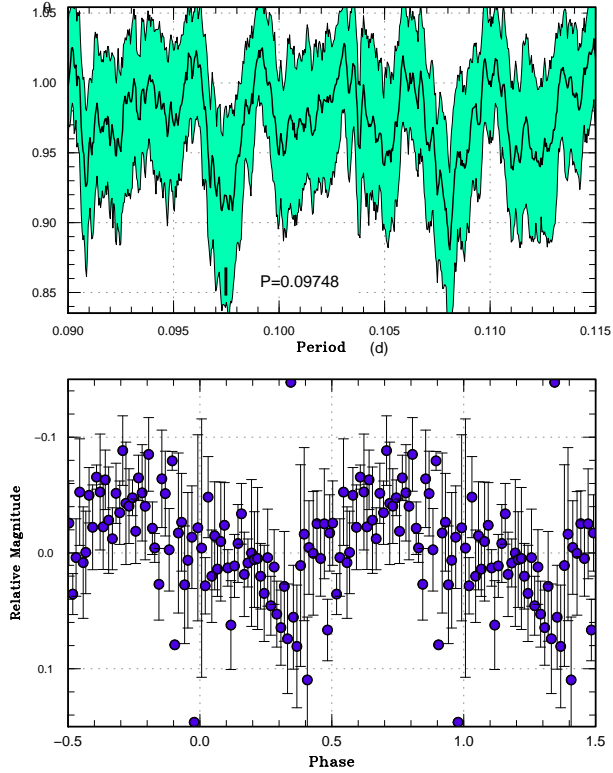}
\end{center}
\caption{Above: PDM analysis for the 2002 first wide outburst
(BJD 2452445--2452448). The 90$\%$ confidence
interval for $\theta$ is shown by green strip. The preferable period is
marked. Below: Phase-averaged profile
of the data folded on the 0.09748~d period.
For clarity data are reproduced twice.}
\label{fig:w2002}
\end{figure}

We selected the five light curves for the 2002
(BJD 2452445--2452449) first wide outburst.
The corresponding PDM
analysis for these data (the central part of the outburst)
is presented in figure \ref{fig:w2002}.  For two
of the most significant 1-day alias signals on the periodogram,
one signal points to the 0.09748(17)~d period that coincides with
the orbital one within the limits of errors.
The mean amplitude of the light curve is $\sim$0.1~mag.
Its profile looks like a round maximum with some depression at
phase 0.9 and with a sharp minimum.
Note that 1-day alias signal points to the 0.108~d period and does not
fall in the range between 0.104~d and 0.106~d that corresponds to
the different estimates of the superhump period according to
\citet{nog98nyser}; \citet{pat03suumas}.

\subsection{2002 Second Wide Outburst}

\begin{figure}
\begin{center}
\FigureFile(80mm,120mm){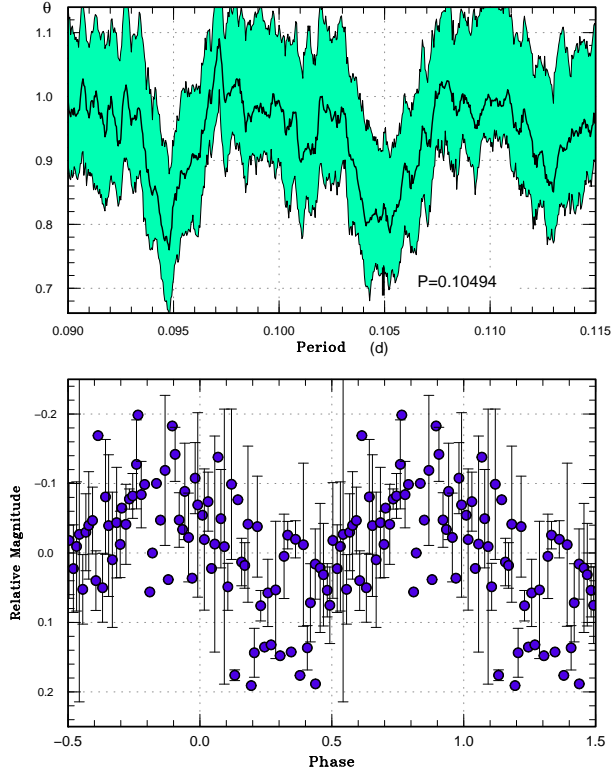}
\end{center}
\caption{Above: PDM for the 2002 second wide outburst superoutburst
(BJD 2452495--BJD 2452499). The 90$\%$ confidence
interval for $\theta$ is shown by green strip. The preferable period is
marked. Below: The average light
curve for the 0.1049~d period. For clarity data are reproduced twice. }
\label{fig:sup2002}
\end{figure}

A PDM analysis for the data
BJD 2452495--2452499 yielded a period which
differs from the previous one (see figure \ref{fig:sup2002}).
One of the 1-day aliased periods 0.10495(13)~d, as in the case
above, also falls in a range of the published superhump periods
(\cite{nog98nyser}; \cite{pat03suumas}).
Since the 1-day alias to the 0.10495(13)~d period is far from the
orbital one, so we accept the 0.10495(13)~d period as
the true one.  The phase-averaged light curve is singly humped and
has an amplitude $\sim$0.2 mag.

We identified that the first wide outburst having
duration 5-6~d is a wide normal outburst, while the second one
represents a fragment of a superoutburst, where the first
three nights probably belong to the superoutburst precursor.

\subsection{2013 First Wide Outburst}

\begin{figure}
\begin{center}
\FigureFile(80mm,120mm){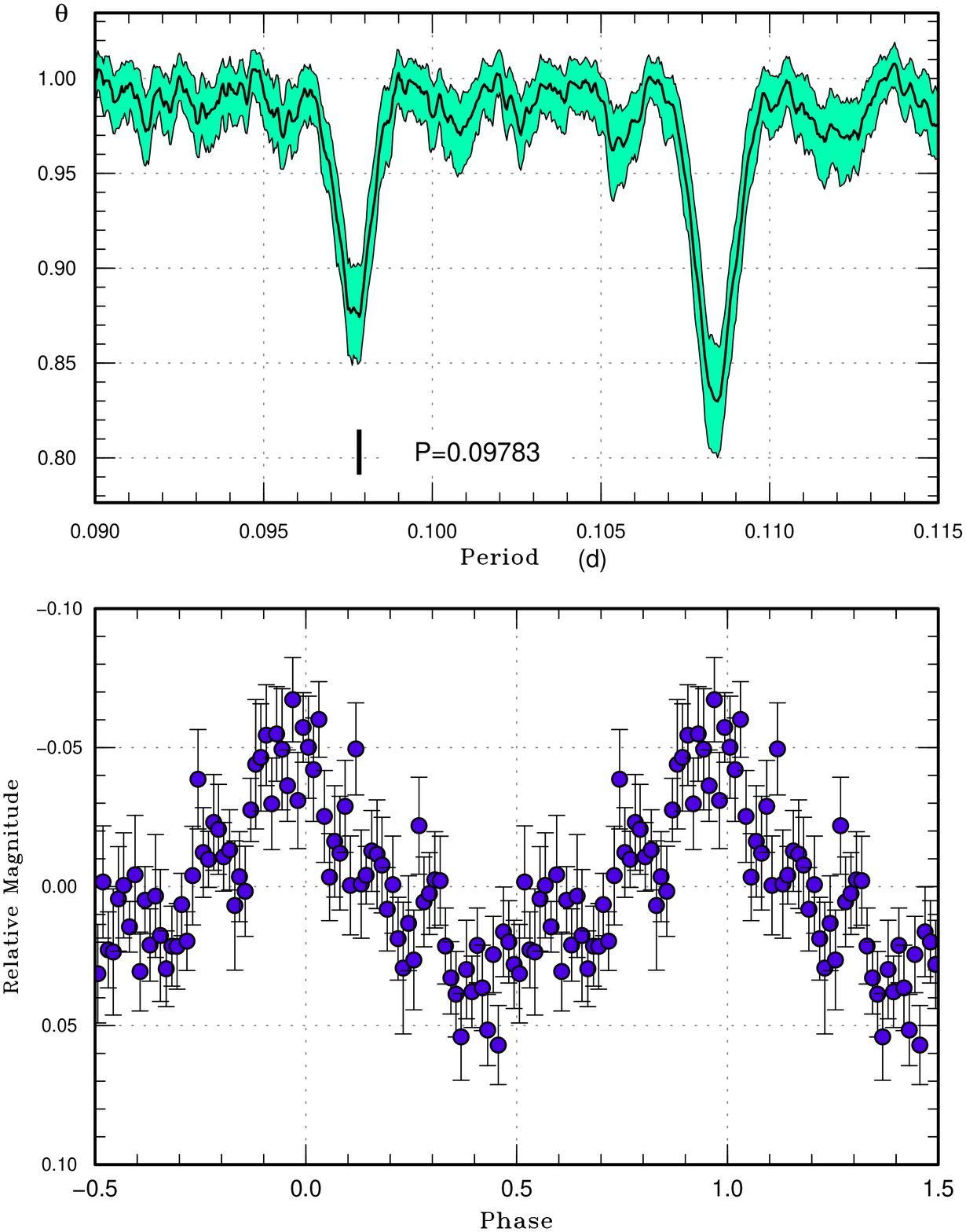}
\end{center}
\caption{Above: PDM analysis for the 2013  first wide outburst.
The 90$\%$ confidence
interval for $\theta$ is shown by green strip. The preferable period is
marked. Below: The phase-averaged light curve for the 0.09783~d period.
For clarity data are reproduced twice.}
\label{fig:w}
\end{figure}

The periodogram obtained with PDM
analysis for all of the data using the most densely observed
part of this outburst (BJD 2456466--2456474) is shown
in figure \ref{fig:w}.
Contrary to our expectation to detect superhumps for this wide outburst,
the periodogram yield the only period
of 0.09783(2)~d that coincides with the orbital one. So we classified
this outburst as the ``wide normal outburst".

\subsection{2013 Second Wide Outburst}

\begin{figure}
\begin{center}
\FigureFile(80mm,120mm){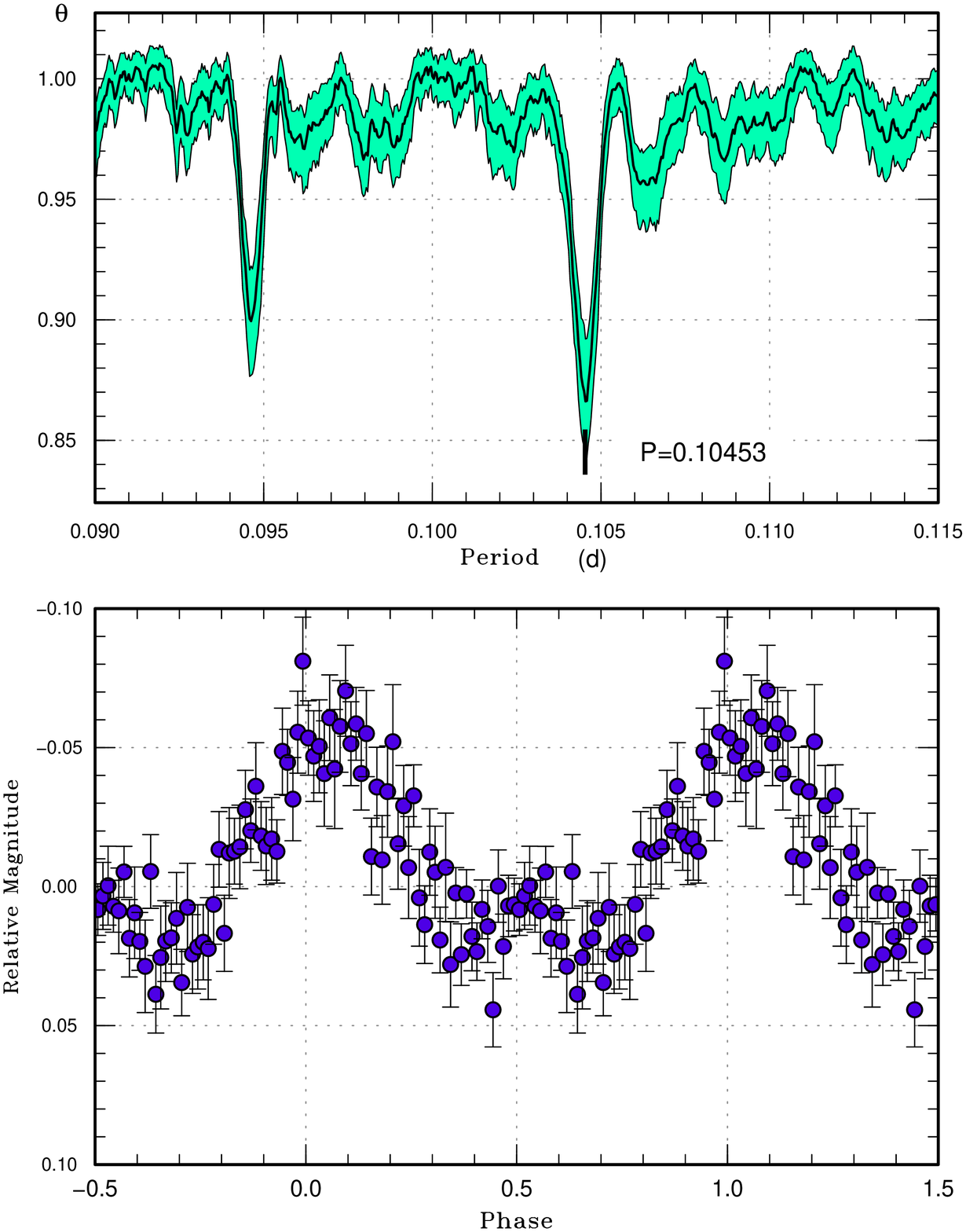}
\end{center}
\caption{Above: PDM analysis fore the 2013 second wide outburst (superoutburst).
 The 90$\%$ confidence interval for $\theta$ is shown by green strip.
The preferable period is
marked. Below: the phase-averaged light curve for the formally best
period of 0.10453~d period.
For clarity data are reproduced twice.}
\label{fig:s2013}
\end{figure}

We selected the data of 12 nights in a region of
BJD 2456489--2456502 for the next wide outburst
followed by the 5-d quiescence after the previous 12-d outburst,
skipping the data of first two nights.
The result of PDM analysis for detrended data
is shown in figure \ref{fig:s2013}.

The most significant signal points to the period
of 0.104531(37)~d.  This period is close to the positive superhump
period obtained for the 2002 superoutburst and to
its evaluations reported by \citet{nog98nyser} and \citet{pat03suumas}.
The mean light curve obtained for this period has singly humped
profile with $\sim$0.1 mag amplitude.
This 18-d wide outburst is identified as a superoutburst.

Using the frequency of this positive superhumps $F_{\rm sh}$
and the frequency of the known orbital period $F_{\rm orb}$, we
obtain the fractional period excess (in the frequency unit)
$\epsilon^* = 1 - F_{\rm sh}/F_{\rm orb} = 0.072$.

\subsection{Periodicity Outside Superoutburst}

In figure \ref{fig:ztsh1}, selected nightly light curves
for quiescent state of NY Ser are shown.
One could see the strong brightness variations on a scale
of $\sim$0.1~d and some low-amplitude oscillations superposed on
the light curves.
Despite of the same mean brightness, the amplitude of the
quiescent light curves is not stable.  Thus it reaches 0.4 mag
on BJD 2456461 and BJD 2456482, but was only $\sim$0.2 mag
on BJD 2456483.  Prominent flickering
with amplitudes up to 0.1--0.2 mag was superposed on some
quiescent light curve.

The figure \ref{fig:ztsh2} displays the nightly light curves
for the eighth normal
outburst in figure \ref{fig:nyser2013} when it approached
 from the maximum to quiescence.  In magnitude unit,
the amplitude of brightness
modulation gradually increases from $\sim$0.05 mag to
$\sim$0.4~mag over the outburst decline.
The shape of the light curves during the top of the outburst and
the middle of decline is rather structured.

\begin{figure}
\begin{center}
\FigureFile(80mm,120mm){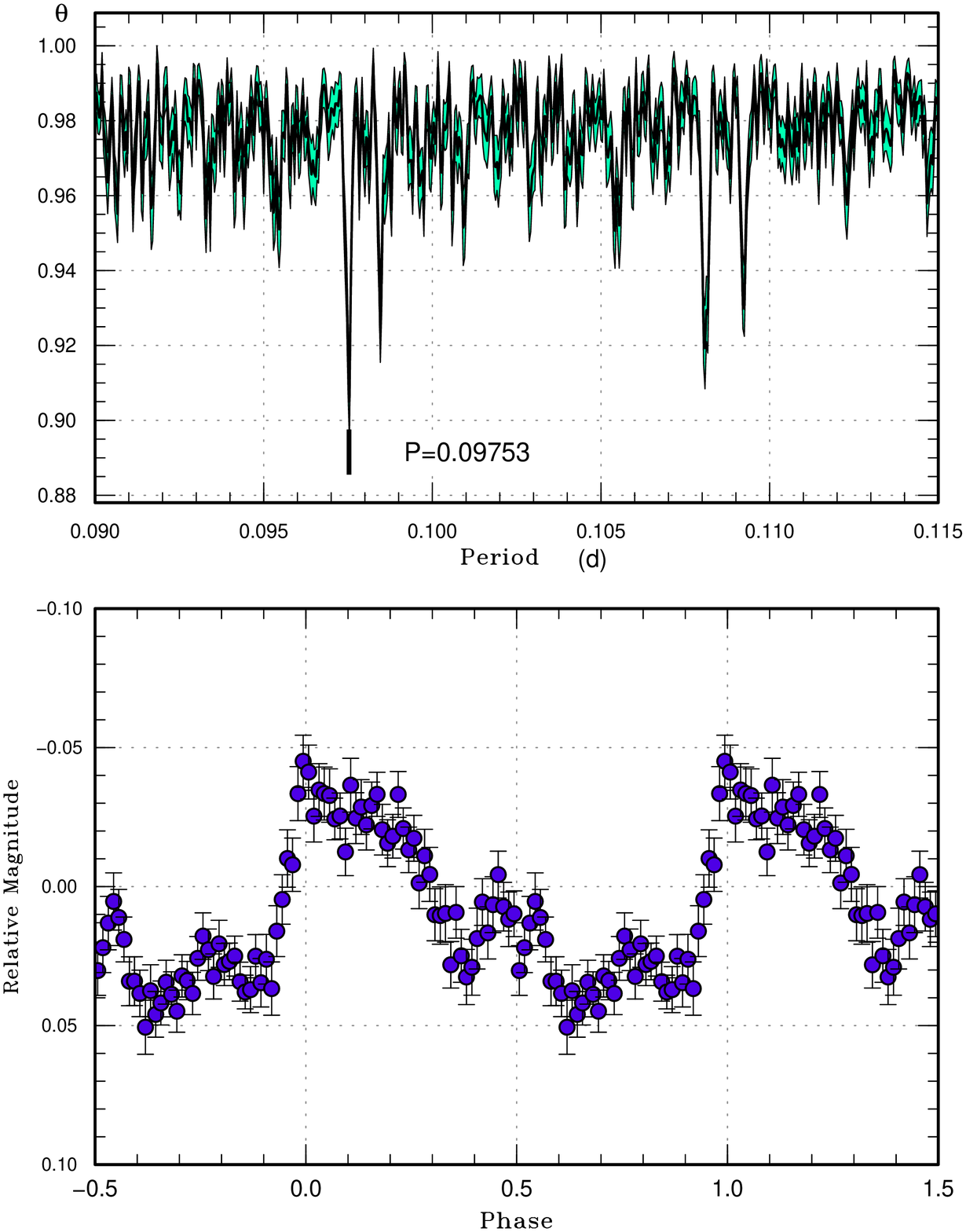}
\end{center}
\caption{Above: PDM analysis for the all data before 2013 superoutburst.
The 90$\%$ confidence
interval for $\theta$ is shown by green strip. The preferable period is
marked. Below: phase-averaged  light curve
for the 0.09753~d period. For clarity data are reproduced twice.}
\label{fig:min1}
\end{figure}

We searched for periodicity of the observations in 2013 of NY Ser 
in the state outside the superoutburst. 
Taking into account that before the superoutburst NY Ser was
in a state of a less frequent
normal outbursts than after the superoutburst,
we studied the periodicity for these states separately.
The first data set included all the data both in minimum
and all normal outbursts before the superoutburst (see figure \ref{fig:min1}).
One could see a sharp
strong signal on the periodogram for the first data set
pointing to the period of 0.097531(4)~d that coincides with
the orbital one.  One-day alias period at 0.108~d
is also presented.  The averaged light curve has the amplitude of
$\sim$1 mag and some dip at a phase around 0.3--0.4.

\begin{figure}
\begin{center}
\FigureFile(80mm,120mm){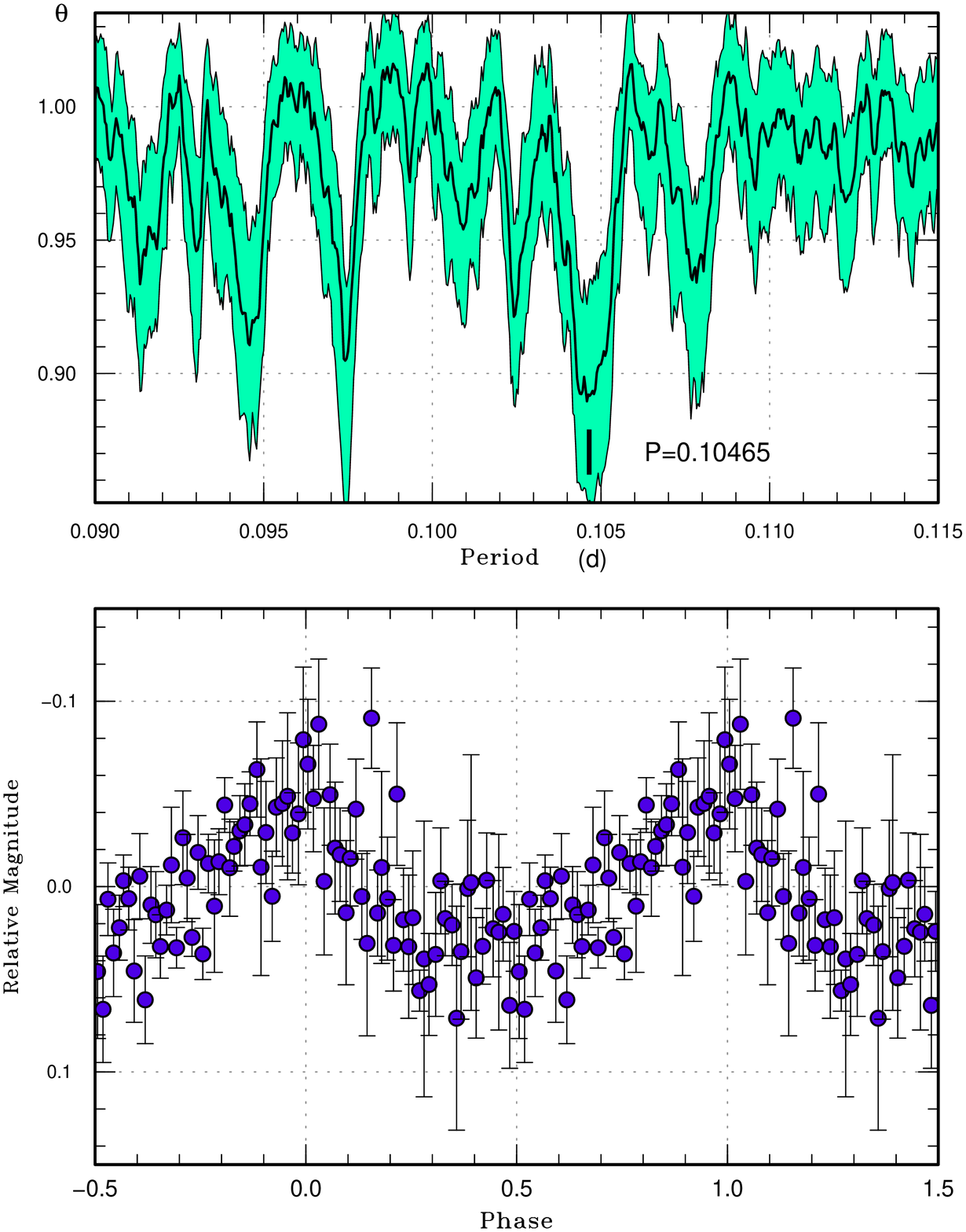}
\end{center}
\caption{Above: PDM analysis fore the data after 2013 superouburst limited
to the fifth and sixth normal outbursts and quiescence around them.
 The 90$\%$ confidence interval for $\theta$ is shown by green strip.
The preferable period is
marked. Below: the phase-averaged light curve for the 0.10465~d period.
 Note that it is contaminated by the orbital modulation.
For clarity data are reproduced twice.}
\label{fig:L}
\end{figure}

The second data set included the data of the highest outburst activity,
i.e., the fifth and sixth normal outbursts and quiescence around them.
The result of PDM analysis is shown in figure \ref{fig:L}.
This periodogram contains the two strongest signals
at the period of 0.10465(12)~d and 0.09744(5)~d, which we believe
to be the period of positive superhumps and the orbital period,
respectively.

Our conclusion is that during the state of relatively infrequent
outbursts the orbital modulation was the dominant signal
both during normal outbursts and in quiescence.
During the state of most frequent outbursts we confidently detected
the co-existence of the surviving positive superhumps and orbital period.

As mentioned above, after this superoutburst the cycle length of
the normal outbursts shortened up to 5~d--6~d.
The effect of a change of the normal outburst frequency was detected
for several stars.  For V1504 Cyg (\cite{osa13v1504cygKepler}) and
V344 Lyr (\cite{osa13v344lyrv1504cyg}) it is established that
the decrease of the frequency of normal outbursts is accompanied
by the negative superhumps appearance.  Vice versa,
the disappearing of the negative superhumps was found for
V503 Cyg [\citet{Pdot4}, \citet{pav12v503cyg}]
during the stage of a frequent normal outbursts.
Since NY Ser entered the stage of frequent
normal outbursts after the 2013 superoutburst,
we checked whether there were negative superhumps or ``impulsive"
negative superhumps [\citet{osa13v344lyrv1504cyg}]
in the data in the state of relatively infrequent normal outbursts
including the wide outburst.

Using the ratio $\epsilon^*_p/\epsilon^*_n \sim 7/4$
[\citet{osa13v344lyrv1504cyg}], where the $\epsilon^*_p$
and the $\epsilon^*_n$ are 0.067 and 0.038, we obtained
the expected period of negative superhumps
to be $\sim$0.09379~d.

 We did not find any indications of negative superhumps
neither for the state with relatively infrequent outburst nor
for the state with frequent outbursts.

\section{Discussions}

\subsection{Variation in Long-Term Behavior and Relation to
Variation in Outburst Activity}

The Catalina Real-time Transient Survey (CRTS; \cite{CRTS})
recorded NY Ser since 2010 May.\footnote{The public data
  is available at
  $<$http://nesssi.cacr.caltech.edu/DataRelease/$>$.
}
The quiescent magnitudes of NY Ser before the 2012 season
were around 18 mag, while they became brighter (17.0--17.5
mag) in the 2012 and 2013 seasons.  This brightening
of the quiescent magnitude may suggest an increase
in the mass-transfer rate or the increased dissipation
in the quiescent disk (e.g. increased quiescent viscosity).
Although the cause of such a variation is not clear,
this systematic variation in the state of this system
may be responsible for the variation in the outburst
activity in the last two years.

\subsection{NY Ser and SU UMa Stars in the Period Gap}

NY Ser is a rare, but not unique, binary among the SU UMa-type
stars showing some deviation from the bimodality of
outbursts.  Yet \citet{war95suuma} (and reference therein) mentioned
that TU Men, the SU UMa dwarf nova with the longest orbital period
within the period gap has trimodal distribution of outburst widths.
The longest outbursts were confirmed to be superoutbursts with
a duration of $\sim$20~d, the wide
outbursts have duration of 8~d and narrow outbursts lasting only
for $\sim$1~d.
Another candidate for an object showing trimodal outbursts is YZ Cnc
[\citet{pat79SH}].  Its period is slightly below the period gap.
NY Ser is the first SU UMa dwarf nova possessing much richer diversity
of the normal outbursts.

In some respect, the morphology of the outbursts of NY Ser
reminds those of CVs above the period gap 
 (having a longer time scale).
Note that the novalikes could
have the same orbital periods as the dwarf novae, but a higher mass transfer rate.
For example, the novalike MV Lyr, located close to the long-period
end of the period gap, at some epochs displayed the variety of
outbursts looking like an alternation of a sequence
``wide outburst --- several narrow outbursts'' without
confirmed superhumps \citep{pav99mvlyr}; \citep{hon04vyscl}.

\begin{figure}
\begin{center}
\FigureFile(80mm,120mm){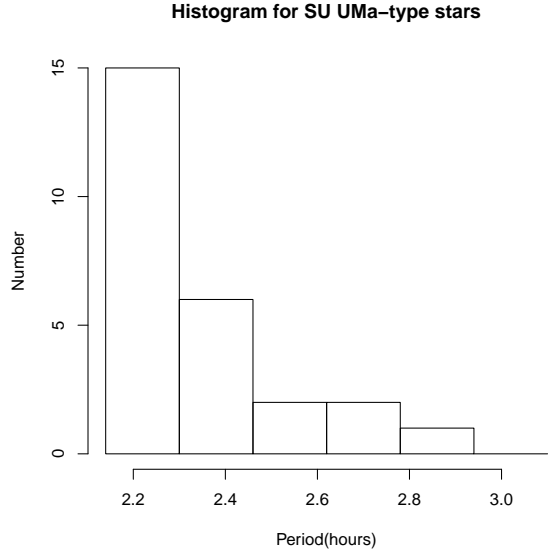}
\end{center}
\caption{Distribution of the orbital periods of the SU UMa-type stars
in the period gap.}
\label{fig:hist}
\end{figure}

The existence of the non-magnetic CVs in the period gap is
still an open question.
According to the theory of magnetic braking \citep{ham88CVmagneticbraking},
when the secondary becomes fully convective,
it shrinks and goes within the Roche lobe that happens at
$\sim$3 hr and fills in it again
at $\sim$2 hr.  According to this most simplest evolutionary scenario,
there should be no CV in the period gap.
However, according to the Ritter and Kolb catalog
(\citet{RKcat}, 7.20 addition) and orbital periods
from \citet{pav10mndra}, Pavlenko et al. (in preparation) for
1RXS J003828.7+250920 [\citet{Pdot3}],
there are 26 known dwarf novae in the 2.15--3.18~hr period gap
up to the middle of 2013.
They are listed in table \ref{tab:DNs}.
The majority of them, namely 23 of systems, are SU UMa-type stars.
The distribution of the orbital periods of these SU UMa-type stars
is presented in figure \ref{fig:hist}.

\begin{table}
\caption{Dwarf Novae In the Gap.}\label{tab:DNs}
\begin{center}
\begin{tabular}{ccc}
\hline
Object Name&Orb.Per&Type\\
 &(day)\commenta& \\
\hline
TU Mensae & 0.1172 & SU \\
V405 Vulpeculae & 0.113l$+$ &  SU \\
CI Geminorum & 0.11$+$ & SU \\
CS Indi & 0.11 & DN \\
AX Capricorni & 0.109$+$ & SU \\
SDSS J162718.39+120435.0 & 0.104$+$ & SU \\
CSS1 0531:134052+151341 & 0.1021 & DN \\
V1239 Herculis & 0.1000 & SU \\
MN Draconis & 0.100$+$ & SU \\
OGLE J175310.04-292120.6 & 0.100$+$ & SU \\
CSS120813:203938-042908.04 & 0.100$+$ & SU \\
V1006 Cygni & 0.09904 & DN \\
NY Serpentis & 0.0978 & SU \\
DV Scorpii & 0.0950$+$ & SU \\
V444 Pegasi & 0.0947$+$ & SU \\
CSS 110628:142548+151502 & 0.094$+$ & SU \\
V725 Aquilae & 0.0939$+$ & SU \\
1RXS J003828.7+250920 & 0.094511 & SU \\
SBSS 0150+339 & 0.093$+$ & SU \\
CSS 110205 J120053-152620& 0.093$+$ & SU \\
CSS 111004:214738+244554 & 0.0927 & SU \\
AD Mensae & 0.0922 & SU \\
TCP J08461690+3115554 & 0.09138 & SU \\
CSS 080427:131626-151313 & 0.091$+$ & SU \\
V589 Herculis & 0.0905 & SU \\
GV Piscium & 0.090$+$ & SU \\
\hline
  \multicolumn{3}{l}{\commenta $+$ means that the orbital period was}\\
 \multicolumn{3}{l}{
indirectly determined from the positive superhumps} \\

\end{tabular}
\end{center}
\end{table}

One could see that this distribution is not uniform.
The number of systems strongly increases
with shortening of the orbital period.
Little more than half of the SU UMa-type stars are
concentrated between 2.18 and 2.3 hr.
However, this tendency apparently only applies to the
distribution of SU UMa stars.
According to \citet{kni06cvsecondary}
the number of all CVs inside the gap displays the opposite behavior,
namely, it increases towards the longer period.

The most accepted explanation of the existence of CVs in the period gap
is that they are formed in the period gap
(they became contact binaries with the orbital periods just
coinciding with periods in the period gap).
We here present two explanations why SU UMa-type stars are
more concentrated in the shorter periods within the period gap.

If we assume that the SU UMa stars with orbital periods inside
the gap are formed uniformly in period, there stars will evolve within
the period gap displaying dwarf nova-type activity.
It would lead to a higher number of objects in short periods
because both objects formed in shorter and longer periods
contribute to the short-period population while only the objects
formed in longer periods contribute to the long-period population.
This effect would resembles accumulation of SU UMa stars
at the short period boundary during evolution.
This explanation, however, cannot explain why the total
number of CVs is larger in region of the longer period
in the period gap.

The second, more plausible explanation is that
the 3:1 resonance becomes harder to reach for the SU UMa stars
with increasing orbital period.  Considering that
the 3:1 resonance responsible for the superhumps
appearing occurs for $q \leq0.25$
[\citet{whi88tidal}, \citet{hir90SHexcess}, \citet{lub91SHa}],
which condition could be expected to be achieved in dwarf novae
closer to the short bound of the period gap.
It seems that NY Ser is not able to achieve the 3:1 resonance
in some of long outbursts.

\subsection{V1006 Cygni Case}

\begin{figure}
\begin{center}
\FigureFile(80mm,120mm){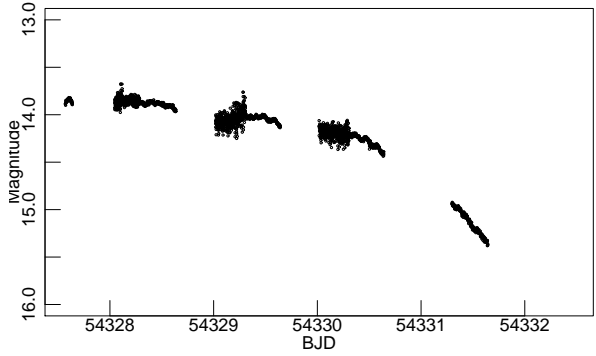}
\end{center}
\caption{The 2007 light curve of V1006 Cyg}
\label{fig:d7}
\end{figure}

\begin{figure}
\begin{center}
\FigureFile(80mm,120mm){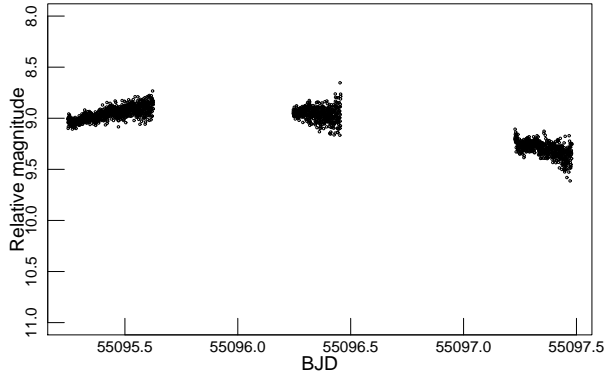}
\end{center}
\caption{The 2009 light curve of V1006 Cyg}
\label{fig:d9}
\end{figure}

According to spectroscopic observations [\citet{she07CVspec}],
the orbital period of V1006 Cyg is 0.09903(9)~d,
qualifying it a dwarf nova in the period gap.
In the RK catalog, V1006 Cyg is classifed as an SU UMa-type star
referring to a preliminary report of a period
(vsnet-alert 9487), which was corrected later
(vsnet-alert 9489).  We here report the result of analysis
of this dwarf nova from by VSNET Collaboration [\citet{VSNET}]
obtained in 2007 and 2009 during its long outbursts.
The light curves of the 2007 and 2009 outbursts are shown
in figures \ref{fig:d7} and \ref{fig:d9}.

\begin{figure}
\begin{center}
\FigureFile(80mm,120mm){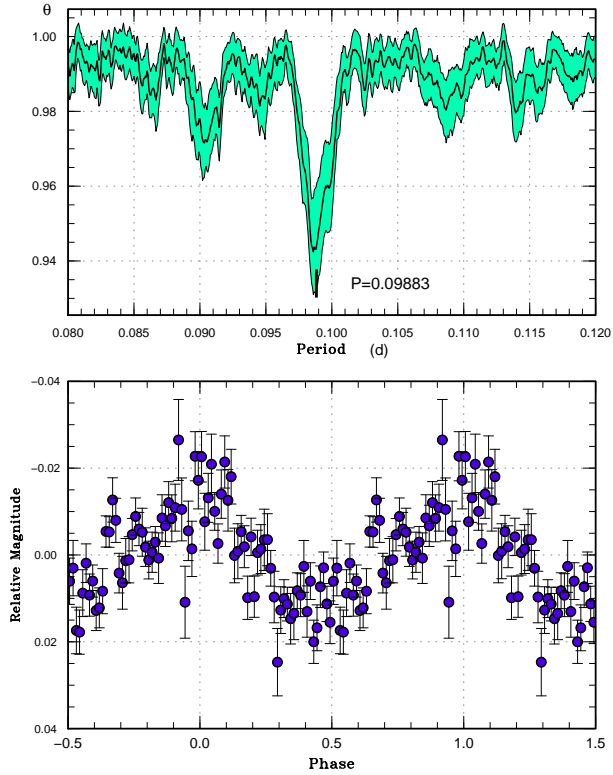}
\end{center}
\caption{Above: PDM analysis for the data of V1006 Cyg (2007 outburst).
The 90$\%$ confidence
interval for $\theta$ is shown by green strip. The preferable period is
marked. Below: phase-averaged  light curve
for the 0.09883~d period. For clarity data are reproduced twice.}
\label{fig:cyg07}
\end{figure}

\begin{figure}
\begin{center}
\FigureFile(80mm,120mm){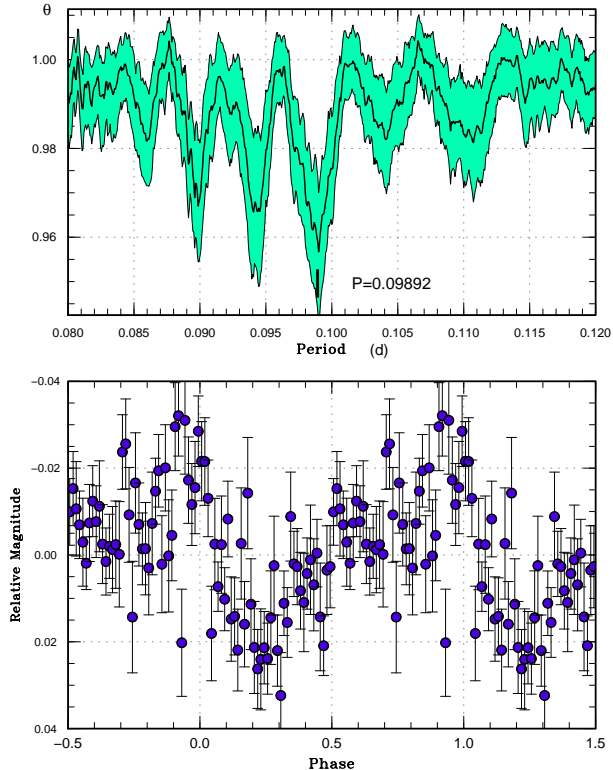}
\end{center}
\caption{Above: PDM analysis for the data of V1006 Cyg (2009 outburst).
The 90$\%$ confidence
interval for $\theta$ is shown by green strip. The preferable period is
marked. Below: phase-averaged  light curve
for the 0.09892~d period. For clarity data are reproduced twice.}
\label{fig:cyg09}
\end{figure}

The PDM analysis yielded the periods 0.09883(12)~d and
0.09892(28)~d for the 2007 and 2009 outbursts respectively
(see figures \ref{fig:cyg07} and \ref{fig:cyg09}), that
within the errors coinside with orbital period discovered
from spectroscopy. The amplitude of light curve for both outbursts
is $\sim$0.02 mag.
So we found only the orbital modulation
in each outburst and
there were no indication to the expected
superhump period.  Since superhumps are not yet detected in
V1006 Cyg, we should regard it as an SS Cyg-type star
rather than an SU UMa-type star.
It is not clear, whether these outbursts represent
the wide ones similar to the NY Ser-like wide outbursts with
orbital modulation, and, similar to NY Ser or TU Men,
one could expect the appearance of superhumps during a much
longer outburst.  On the other hand, V1006 Cyg would remain
a genuine SS Cygni-type star which never reaches the 3:1 resonance.
The presence of an apparent SS Cyg-type star having
long outbursts without superhumps in the period gap
supports our second suggestion that the 3:1 resonance
is harder to achieve in longer period in the period gap
and it would explain the distribution of SU UMa stars
inside the period gap.

\section{Conclusion}

In this work we discovered in NY Ser unusually long 12-d outburst
without superhumps but displaying the orbital light variations.

\citet{war95suuma} in his monograph considering the SU UMa-type stars wrote:
``if DN are ever found with superoutbursts lacking superhumps
they will define a class of their own".
NY Ser could be an example of an intermediate subclass
of the dwarf novae possessing some properties of
the cataclysmic variables below and above the period gap.

During the state of relatively infrequent outbursts
the orbital modulation was the dominant signal both during
normal outbursts and in quiescence.
During the state of most frequent outbursts followed immediately
by the 2013 superoutburst, we confidently detected
the co-existence of the surviving positive superhumps and
orbital period.

We were not able to detect a sufficient number of the superhump maxima
during the superoutbursts because of the limited length of nightly
light curves, we leave for future the study of evolution of
the positive superhumps in this system.
We hope that within the future international
multi-longitude campaign we could also define the frequency of
the appearance of different types of normal outbursts and understand the
phenomenon of this unique dwarf nova that could bring us closer to
understanding the evolution of CVs inside the period gap.

We are grateful for the suggestions from Prof. Yoji Osaki
during the conference ``Kyoto Mini-Workshop on Dwarf Novae and
Related Systems --- New Directions in Time-Series Analysis"
in 2013 October. We thank the anonymous referee whose
comments greatly improved the paper.
This work was supported by the Grant-in-Aid Initiative for
High-Dimensional Data-Driven Science through Deepening of
Sparse Modeling from the Ministry of Education,
Culture, Sports, Science and Technology (MEXT) of Japan.
S. Shugarov
is grateful for support by the RFBR grant 11-02-00258 (Russia) and
VEGA Grant No. 2/0002/13 (Slovakia).
K. Antonyuk and N. Pit express a specific acknowledgment to
the funding of the CCD Camera FLI ProLine PL230 by Labex OSUG@2020.

\begin{table*}
\caption{Observation list }\label{tab:log}
\begin{center}
\begin{tabular}{ccccc}
\hline
JD start\commenta & JD end\commenta & N\commentb & Obs\commentc & Exp\commentd\\
\hline
52435.3263 & 52435.3312 &   2 & B & 360 \\
52439.3575 & 52439.4165 &  14 & B & 300 \\
52441.3552 & 52441.3649 &   3 & B & 240 \\
52442.3446 & 52442.4178 &  25 & B & 300 \\
52445.3051 & 52445.418  & 141 & B & 60 \\
52446.3496 & 52446.4398 &  74 & B & 90 \\
52447.3253 & 52447.4467 &  73 & B & 120 \\
52448.3293 & 52448.4421 &  77 & B & 120 \\
52449.3179 & 52449.4399 &  96 & B & 90 \\
52450.3542 & 52450.4862 &  26 & B & 360 \\
52451.3588 & 52451.5138 &  50 & B & 240 \\
52453.3394 & 52453.4962 &  23 & B & 240 \\
52454.3737 & 52454.5166 &  62 & B & 180 \\
52455.343  & 52455.4829 &  46 & B & 180 \\
52456.3651 & 52456.4819 &  40 & B & 180 \\
52458.3027 & 52458.331  &   9 & B & 240 \\
52482.308  & 52482.3371 &  16 & B & 120 \\
52483.2906 & 52483.319  &  19 & B & 120 \\
52484.2666 & 52484.2766 &   7 & B & 120 \\
52485.3262 & 52485.3316 &   3 & B & 240 \\
52488.2825 & 52488.2878 &   4 & B & 180 \\
52489.2851 & 52489.2903 &   3 & B & 180 \\
52491.3048 & 52491.3438 &  28 & B & 90 \\
52492.2937 & 52492.3248 &  24 & B & 90 \\
52493.2629 & 52493.3079 &  33 & B & 90 \\
52495.3304 & 52495.366  &  18 & B & 120 \\
52496.2717 & 52496.3596 &  44 & B & 120 \\
\hline
\multicolumn{5}{l}{\commenta JD$-$2456100} \\
\multicolumn{5}{l}{\commentb Number of observations} \\
\multicolumn{5}{l}{\commentc A$=$ CrAO/K-380/Apogee E47/Clear;} \\
\multicolumn{5}{l}{B$=$CrAO/K-380/SBIG ST7/Clear; C$=$CrAO/AZT11/FLI1001E/Clear }\\
\multicolumn{5}{l}{D$=$Tatranska Lomnica/Zeiss600/Apogee U9000/Clear;}\\
\multicolumn{5}{l}{E$=$SAI/Zeiss600/versarray 1300/Clear;}\\
\multicolumn{5}{l}{F$=$CrAO/ZTSh/Apogee E47/R; G$=$/Chile40cm/FLI16803 CCD/V}\\
\multicolumn{5}{l}{\commentd Exposure time (sec)}\\
\end{tabular}
\end{center}
\end{table*}
\addtocounter{table}{-1}
\begin{table*}
\caption{Observation list(continued)}
\begin{center}
\begin{tabular}{ccccc}
\hline
JD start\commenta & JD end\commenta & N\commentb & Obs\commentc & Exp\commentd\\
\hline
52497.269  & 52497.3471 &  89 & B & 60 \\
52498.2657 & 52498.3403 &  45 & B & 120 \\
52499.2496 & 52499.3243 &  23 & B & 120 \\
52500.2553 & 52500.2679 &   8 & B & 120 \\
56441.3702 & 56441.4995 & 177 &A & 60 \\
56445.3450 & 56445.4973 & 72 &A & 180 \\
56449.3425 & 56449.4805 & 186 &A & 60 \\
56451.3487 & 56451.5151 & 227 &A & 60 \\
56452.2717 & 56452.5306 & 347 &A & 60 \\
56452.3527 & 56452.5311 & 1038 &E & 30 \\
56453.2942 & 56453.4279 & 63 &A & 180 \\
56454.2817 & 56454.4716 & 75 &A & 180 \\
56454.3144 & 56454.4189 & 73 &E & 60 \\
56455.3054 & 56455.3073 & 3 &E & 60 \\
56455.3216 & 56455.3641 & 21 &A & 180 \\
56457.3087 & 56457.5093 & 307 &C & 60 \\
56461.3803 & 56461.5152 & 350 &F & 30 \\
56461.3883 & 56461.5013 & 55 &C & 180 \\
56462.2901 & 56462.5301 & 139 &A & 120 \\
56463.3134 & 56463.3317 & 5 &A & 180 \\
56466.3195 & 56466.4122 & 66 &A & 120 \\
56467.3698 & 56467.4349 & 46 &A & 120 \\
56469.3442 & 56469.5135 & 232 &A & 60 \\
56470.3058 & 56470.5177 & 287 &A & 60 \\
56471.4078 & 56471.4791 & 98 &A & 60 \\
56472.3177 & 56472.3954 & 166 &A & 30 \\
56473.2826 & 56473.4284 & 199 &A & 60 \\
\hline
\multicolumn{5}{l}{\commenta JD$-$2456100} \\
\multicolumn{5}{l}{\commentb Number of observations} \\
\multicolumn{5}{l}{\commentc A$=$ CrAO/K-380/Apogee E47/Clear;} \\
\multicolumn{5}{l}{B$=$CrAO/K-380/SBIG ST7/Clear; C$=$CrAO/AZT11/FLI1001E/Clear }\\
\multicolumn{5}{l}{D$=$Tatranska Lomnica/Zeiss600/Apogee U9000/Clear;}\\
\multicolumn{5}{l}{E$=$SAI/Zeiss600/versarray 1300/Clear;}\\
\multicolumn{5}{l}{F$=$CrAO/ZTSh/Apogee E47/R; G$=$/Chile40cm/FLI16803 CCD/V}\\
\multicolumn{5}{l}{\commentd Exposure time (sec)}\\
\end{tabular}
\end{center}
\end{table*}
\addtocounter{table}{-1}
\begin{table*}
\caption{Observation list(continued)}
\begin{center}
\begin{tabular}{ccccc}
\hline
JD start\commenta & JD end\commenta & N\commentb & Obs\commentc & Exp\commentd\\
\hline
56473.3140 & 56473.4625 & 208 &A & 60 \\
56474.3679 & 56474.4392 & 108 &A & 60 \\
56475.3226 & 56475.3369 & 3 &D & 160 \\
56476.3648 & 56476.3785 & 7 &D & 180 \\
56477.3357 & 56477.3416 & 6 &D & 300 \\
56478.3342 & 56478.3399 & 6 &D & 60 \\
56479.2999 & 56479.4129 & 55 &C & 180 \\
56480.2896 & 56480.4382 & 72 &C & 180 \\
56481.3407 & 56481.4181 & 38 &C & 180 \\
56482.2958 & 56482.5008 & 764 &F & 20 \\
56482.3603 & 56482.3649 & 6 &D & 60 \\
56483.2875 & 56483.4152 & 62 &E & 180 \\
56483.3296 & 56483.5040 & 647 &F & 20 \\
56483.3402 & 56483.3467 & 7 &D & 60 \\
56484.3876 & 56484.4982 & 509 &F & 20 \\
56485.2986 & 56485.3112 & 5 &A & 180 \\
56486.4483 & 56486.4922 & 62 &C & 60 \\
56489.2660 & 56489.4673 & 270 &C & 60 \\
56490.3873 & 56490.4086 & 11 &A & 180 \\
56491.4279 & 56491.4343 & 4 &A & 180 \\
56492.2856 & 56492.3301 & 22 &A & 180 \\
56492.3073 & 56492.4290 & 173 & C & 60 \\
56493.2781 & 56493.2918 & 7 &A & 180 \\
56493.3534 & 56493.4731 & 165 &C & 60 \\
56494.2734 & 56494.3625 & 15 &A & 360 \\
56495.3015 & 56495.3778 & 13 &A & 360 \\
56496.2693 & 56496.3866 & 32 &A & 300 \\
\hline
\multicolumn{5}{l}{\commenta JD$-$2456100} \\
\multicolumn{5}{l}{\commentb Number of observations} \\
\multicolumn{5}{l}{\commentc A$=$ CrAO/K-380/Apogee E47/Clear;} \\
\multicolumn{5}{l}{B$=$CrAO/K-380/SBIG ST7/Clear; C$=$CrAO/AZT11/FLI1001E/Clear }\\
\multicolumn{5}{l}{D$=$Tatranska Lomnica/Zeiss600/Apogee U9000/Clear;}\\
\multicolumn{5}{l}{E$=$SAI/Zeiss600/versarray 1300/Clear;}\\
\multicolumn{5}{l}{F$=$CrAO/ZTSh/Apogee E47/R; G$=$/Chile40cm/FLI16803 CCD/V}\\
\multicolumn{5}{l}{\commentd Exposure time (sec)}\\
\end{tabular}
\end{center}
\end{table*}
\addtocounter{table}{-1}
\begin{table*}
\caption{Observation list(continued)}
\begin{center}
\begin{tabular}{ccccc}
\hline
JD start\commenta & JD end\commenta & N\commentb & Obs\commentc & Exp\commentd\\
\hline
56496.5271 & 56496.5957 & 33 &G & 180 \\
56497.2730& 56497.4430& 233 &A & 240 \\
56497.4702& 56497.5924& 61 &G & 160 \\
56498.3073& 56498.4499& 60 &A & 120 \\
56498.3156& 56498.4423& 176 &C & 120 \\
56498.4704& 56498.5887& 82 & G & 160 \\
56499.3218& 56499.4159& 48 &A & 180 \\
56499.3272& 56499.3371& 15 &C & 60 \\
56499.5476& 56499.5858& 19 &G & 160 \\
56500.2790& 56500.3809& 48 &A & 180 \\
56500.3314& 56500.4179& 121 &C & 60 \\
56501.3240& 56501.3326&   6 & A & 120 \\
56501.4708& 56501.5810& 52 &G & 180 \\
56502.3667& 56502.3999&  16 & A & 180 \\
56502.4716& 56502.5799& 53 &G & 170 \\
56503.2984& 56503.3153&   7 & A & 180 \\
56503.4713& 56503.5754& 50 &G & 170 \\
56504.3138& 56504.4283&  53 & A & 180 \\
56505.2762& 56505.3313&  27 & A & 180 \\
56505.2959& 56505.3320& 51 &C & 60 \\
56505.5348& 56505.5710& 19 &G & 170 \\
56506.3418& 56506.3609&  10 & A & 180 \\
56506.4719& 56506.5688& 47 &G & 170 \\
56507.2924& 56507.3698&  24 & A & 120 \\
56507.3073& 56507.3828& 55 &C & 120 \\
56507.4721& 56507.5649& 45 &G & 170 \\
56508.3065& 56508.4126&  51 & A & 180 \\
\hline
\multicolumn{5}{l}{\commenta JD$-$2456100} \\
\multicolumn{5}{l}{\commentb Number of observations} \\
\multicolumn{5}{l}{\commentc A$=$ CrAO/K-380/Apogee E47/Clear;} \\
\multicolumn{5}{l}{B$=$CrAO/K-380/SBIG ST7/Clear; C$=$CrAO/AZT11/FLI1001E/Clear }\\
\multicolumn{5}{l}{D$=$Tatranska Lomnica/Zeiss600/Apogee U9000/Clear;}\\
\multicolumn{5}{l}{E$=$SAI/Zeiss600/versarray 1300/Clear;}\\
\multicolumn{5}{l}{F$=$CrAO/ZTSh/Apogee E47/R; G$=$/Chile40cm/FLI16803 CCD/V}\\
\multicolumn{5}{l}{\commentd Exposure time (sec)}\\
\end{tabular}
\end{center}
\end{table*}
\addtocounter{table}{-1}
\begin{table*}
\caption{Observation list(continued)}
\begin{center}
\begin{tabular}{ccccc}
\hline
JD start\commenta & JD end\commenta & N\commentb & Obs\commentc & Exp\commentd\\
\hline
56508.4721& 56508.5628& 44 &G & 170 \\
56509.2565& 56509.3647&  41 & A & 180 \\
56510.3063& 56510.4042&  71 & C & 120 \\
56510.3064& 56510.4043& 71 &A & 180 \\
56511.2908& 56511.3705& 58 &C & 120 \\
56512.4752& 56512.5492& 36 &G & 170 \\
56515.3177& 56515.3223&  16 & A & 120 \\
56516.2508& 56516.3989&  69 & A & 180 \\
56517.2959& 56517.3616&  32 & A  & 180 \\
56518.2611& 56518.3778&  46 & A & 180 \\
56519.2503& 56519.3904&  66 & A & 180 \\
56520.3158& 56520.3879&  34 & A & 180 \\
56521.2745& 56521.3823& 393 &F & 20 \\
56522.2437& 56522.3745& 480 &F & 20 \\
56523.2494& 56523.3608& 207 &F & 20 \\
56524.2486& 56524.2888&  19 & A & 180 \\
56524.2671& 56524.3205& 31 &C & 180 \\
56525.2406& 56525.3445&  50 & A & 180 \\
56526.2532& 56526.3596&  51 & A & 180 \\
56526.2629& 56526.3456& 39 &C & 180 \\
56527.3087& 56527.3450&  18 & A & 180 \\
56528.2516& 56528.3321&  16 & A & 180 \\
56528.2811& 56528.3447& 31 &C & 180 \\
56530.2516& 56530.3258& 36 &C & 180 \\
56530.2784& 56530.3399&  30 & A & 180 \\
56531.2930& 56531.3502&  28 & A & 180 \\
56532.2522& 56532.3158& 31 &C & 180 \\
\hline
\multicolumn{5}{l}{\commenta JD$-$2456100} \\
\multicolumn{5}{l}{\commentb Number of observations} \\
\multicolumn{5}{l}{\commentc A$=$ CrAO/K-380/Apogee E47/Clear;} \\
\multicolumn{5}{l}{B$=$CrAO/K-380/SBIG ST7/Clear; C$=$CrAO/AZT11/FLI1001E/Clear }\\
\multicolumn{5}{l}{D$=$Tatranska Lomnica/Zeiss600/Apogee U9000/Clear;}\\
\multicolumn{5}{l}{E$=$SAI/Zeiss600/versarray 1300/Clear;}\\
\multicolumn{5}{l}{F$=$CrAO/ZTSh/Apogee E47/R; G$=$/Chile40cm/FLI16803 CCD/V}\\
\multicolumn{5}{l}{\commentd Exposure time (sec)}\\
\end{tabular}
\end{center}
\end{table*}
\addtocounter{table}{-1}
\begin{table*}
\caption{Observation list(continued)}
\begin{center}
\begin{tabular}{ccccc}
\hline
JD start\commenta & JD end\commenta & N\commentb & Obs\commentc & Exp\commentd\\
\hline
56533.2402& 56533.3505&  53 & A & 180 \\
56534.2490& 56534.2533& 3 &C & 180 \\
56536.2947& 56536.3002&  18 & A & 180 \\
56537.2451& 56537.2535& 5 &C & 180 \\
\hline
\multicolumn{5}{l}{\commenta JD$-$2456100} \\
\multicolumn{5}{l}{\commentb Number of observations} \\
\multicolumn{5}{l}{\commentc A$=$ CrAO/K-380/Apogee E47/Clear;} \\
\multicolumn{5}{l}{B$=$CrAO/K-380/SBIG ST7/Clear;C$=$CrAO/AZT11/FLI1001E/Clear}\\
\multicolumn{5}{l}{D$=$Tatranska Lomnica/Zeiss600/Apogee U9000/Clear;}\\
\multicolumn{5}{l}{E$=$SAI/Zeiss600/versarray 1300/Clear;}\\
\multicolumn{5}{l}{F$=$CrAO/ZTSh/Apogee E47/R;G$=$/Chile40cm/FLI16803 CCD/V}\\
\multicolumn{5}{l}{\commentd Exposure time (sec)}\\
\end{tabular}
\end{center}
\end{table*}

\newcommand{\noop}[1]{}


\end{document}